\documentstyle [12pt,epsf,rotate] {article}
\newcommand{\nwc}{\newcommand}
\nwc{\be}  {\begin{equation}}
\nwc{\ee}  {\end{equation}}
\nwc{\ba}  {\begin{array}}
\nwc{\ea}  {\end{array}}
\nwc{\bdm} {\begin{displaymath}}
\nwc{\edm} {\end{displaymath}}
\nwc{\bea} {\be\ba{rcl}}
\nwc{\eea} {\ea\ee}
\nwc{\bear} {\begin{eqnarray}}
\nwc{\ear} {\end{eqnarray}}

\begin{document}
\begin{titlepage}
\begin{flushright}
HD--THEP--99--16
\end{flushright}
\quad\\
\vspace{1.8cm}

\begin{center}
{\Large Gluon-Meson Duality}\\
\vspace{2cm}
Christof Wetterich\footnote{e-mail: C.Wetterich@thphys.uni-heidelberg.de}\\
\bigskip
Institut  f\"ur Theoretische Physik\\
Universit\"at Heidelberg\\
Philosophenweg 16, D-69120 Heidelberg\\
\vspace{3cm}

\end{center}

\begin{abstract}
The QCD-vacuum is characterized by the Higgs phenomenon for colored
scalar fields. In this dual picture the gluons appear as the octet of vector
mesons. Also quarks and baryons are identified. Gluon-meson and quark-baryon
duality can account in a simple way for realistic masses of all low-mass
hadrons and for their interactions.

\end{abstract}
\end{titlepage}

\underbar{Universality} is a central concept in today's understanding
of how renormalizable theories arise from unknown ``microscopic physics''
at distances much smaller than directly observable by experiment.
The precise form of the microscopic action does not matter -- a fact
extensively used in lattice simulations of QCD. For example, we may add
to the microscopic Lagrangian of QCD a four-quark interaction
\be\label{A}
\Delta{\cal L}=h^2\left(\bar\Psi_{ai}\left(\frac{1-\gamma_5}
{2}\right)\Psi_{ib}\right)(m^2_\phi-\partial^\mu
\partial_\mu)^{-1}\left(\bar\Psi_{bj}\frac{1+\gamma_5}{2}\Psi_{ja}
\right)\ee
without altering the predictions at observable distances. (Here $i,j$
and $a,b$ are color and flavor indices.) At the microscopic scale
the parameters $h$ and $m_\phi$ are arbitrary as long as $m_\phi$
is not much smaller than the inverse microscopic length scale
$\Lambda$, such that $\Delta{\cal L}$ remains essentially local.
In contrast, effective couplings of the type (\ref{A}) at some scale
$k\ll\Lambda$ become calculable in terms of the renormalized gauge
coupling.

\underbar{Field representations} are not a universal feature. In fact,
we may interpret $\Delta{\cal L}$ as arising from integrating out a
heavy scalar field with Yukawa interaction $h$ to the quarks. In
principle, the choice of fields used to describe a given
universality class is a matter of convenience, provided they account
for the light physical particles and the symmetries governing
their interactions.

\underbar{Duality} is a powerful tool which can be used to relate
different field descriptions for a given universality class. Each description
may be optimal for a given momentum scale $k$ or a given size of the relevant
coupling constants. For example, the nonlinear $O(3)$-Heisenberg model
in two dimensions is asymptotically free. At long distances the dimensionless
coupling grows and produces a non-perturbative mass by dimensional
transmutation. By virtue of these similarities this model has
often been considered as a prototype for QCD. In the non-linear
$\sigma$-model description the nonperturbative behavior for large
coupling appears to be quite complicated. In contrast, a dual
description in terms of a linear $\sigma$-model has provided \cite{1}
a very simple successful picture of the strong coupling region --
it is simply given by a massive $O(3)$-symmetric scalar field
theory. For QCD, we would like to have a similar simple
dual description for scales where the gauge coupling is large. We
already know the relevant degrees of freedom from experiment: For
bosons these are the light mesons and we have to search for
gluon-meson duality. For fermions the light degrees of freedom
are the baryon octet -- we therefore also need quark-baryon duality.
In this note we propose such a simple dual description by using
additional scalar fields. The essential features of confinement
are described by the Higgs phenomenon for colored scalar fields,
which turns the gluons into massive vector mesons. The equivalence
of the confinement or Higgs descriptions has, in principle,
been known for a long time \cite{2}. It was recently used in order
to establish that there is no phase transition in the high temperature
electroweak standard model \cite{3} and for a discussion
of properties of high density quark or nuclear matter \cite{4}.

In addition to the quark and gluon fields we consider scalar
fields with the transformation properties of
quark-antiquark pairs. With respect to the color and chiral flavor
rotations $SU(3)_C\times SU(3)_L\times SU(3)_R$ the three
light left-handed and right-handed quarks $\Psi_L,\Psi_R$
transform as (3,3,1) and (3,1,3), respectively. Quark-antiquark
bilinears therefore contain a color singlet $\Phi(1,\bar 3, 3)$ and
a color octet $\chi(8,\bar 3, 3)$
\bear\label{2}
\gamma_{ij}&=&\chi_{ij}+\frac{1}{\sqrt3}\phi,\nonumber\\
\phi&=&\frac{1}{\sqrt3}\gamma_{ii}\quad,\quad \chi_{ii}=0\ear
Here we use a matrix notation for the flavor indices and write
the color indices explicitly, e.g. $\Psi_{ai}\equiv\Psi_i,\ \chi
_{ij,ab}\equiv\chi_{ij},\ \phi_{ab}
\equiv\phi$. Then $\gamma_{ij}$ contains 81 complex scalar fields.
We consider an effective Lagrangian with terms up to dimension
four
\bear\label{3}
{\cal L}&=&i\bar\Psi_i\gamma^\mu\partial_\mu\Psi_i+g\bar\Psi_i
\gamma^\mu A_{ij,\mu}\Psi_j+\frac{1}{2}G^{\mu\nu}_{ij}
G_{ij,\mu\nu}\nonumber\\
&&+Tr\{(D^\mu\gamma_{ij})^\dagger D_\mu \gamma_{ij}\}+U(\gamma)
\nonumber\\
&&+\bar\Psi_i[(h\phi\delta_{ij}+\tilde h\chi_{ij})
\frac{1+\gamma_5}{2}-(h\phi^\dagger\delta_{ij}+\tilde h
\chi^\dagger_{ji})\frac{1-\gamma_5}{2}]\Psi_j\ear
with $A_{ij,\mu}=\frac{1}{2}A^z_\mu(\lambda_z)_{ij},
\ G_{ij,\mu\nu}=\partial_\mu A_{ij,\nu}-\partial_\nu A_{ij,\mu}-ig
A_{ik,\mu}A_{kj,\nu}+ig A_{ik,\nu}A_{kj,\mu}$ and
$\lambda_z$ the eight Gell-Mann matrices, $Tr\ \lambda_y\lambda_z
=2\delta_{yz}$. The interaction between gluons and $\chi$ arise
from the covariant derivative $D_\mu\gamma_{ij}=\partial_\mu\gamma_{ij}-
ig A_{ik,\mu}\gamma_{kj}+ig\gamma_{ik}A_{kj,\mu}$. In our notation the
transposition acts only on flavor indices, e.g. $(\gamma^\dagger
_{ij})_{ab}=\gamma^*_{ij,ba}$.
The effective potential
\be\label{3a}
U(\gamma)=U_0(\chi,\phi)-\frac{1}{2}\nu (\det\ \phi+\det \phi^\dagger)\ee
conserves axial $U(1)$ symmetry except for the term $\sim\nu$.
(We omit for simplicity other possible $U(1)_A$-violation
by terms involving $\chi$.)
Finally, explicit chiral symmetry breaking is induced
by a linear term
\bear\label{3b}
{\cal L}_j&=&-\frac{1}{2}Z^{-1/2}_\phi\ Tr\ (j^\dagger\phi+\phi^\dagger  
j)\nonumber\\
j^\dagger&=&a_q\  diag(\bar m_u,\bar m_d,\bar m_s).\ear

In a renormalization group framework the parameters appearing in (\ref{2})
should be considered as running coupling constants. For instance,
we may associate  $\Gamma_k=\int d^4x{\cal L}$ with the effective
average action \cite{5} which exhibits an infrared cutoff $k$. The vacuum
properties and the physical particle spectrum should be extracted
for $k=0$. For short-distance scattering processes, however, one
may account for the momentum dependence of the vertices by
using for $k$ an appropriate momentum scale. Our description should
coincide with perturbative QCD for large $k$.

Let us expand $U_0=m^2_\phi\ Tr\ \phi^\dagger\phi+m^2_\chi\ Tr\ \chi
^\dagger_{ij}\chi_{ij}+...$ and assume that for large $k$
both $m_\phi^2$ and $m^2_\chi$ are large and positive. Then
the additional scalar excitations will effectively decouple. Whereas
the expectation value $<\chi>$ vanishes, one finds
$<\phi>=\frac{1}{2}m^{-2}_\phi Z^{-1/2}_\Phi j$ and therefore
effective quark masses
\be\label{3c}
m_q=\frac{1}{2}hm_\Phi^{-2}Z_\Phi^{-1/2}j_q.\ee
This is the only relevant effect of the additional degrees
of freedom and exactly what is needed for perturbative
QCD! We have explicitly checked that in leading order the
running of the quark masses only arises from gluon diagrams.
(The sources $j$ are constant and the effects of the
running scalar wave function renormalization $Z_\Phi$ drop out.)
In fact, for large $m^2_\Phi$ and $m^2_\chi$
it is straightforward to integrate out the scalar degrees of freedom.
The field $\Phi$ produces a correction
$\Delta{\cal L}$ given by (\ref{A}) and the quark mass term
(\ref{3c}), whereas from $\chi$ we obtain higher-order gluon
interactions by expanding $-Tr\ \ln(m^2_\chi-D^2[A])$ with $D^2$
the covariant Laplacian and a term similar to (\ref{A}).
The corrections are
nonrenormalizable interactions which
are irrelevant by standard universality arguments. In fact, the
scalar fields in $\Gamma_k$ may be viewed as a shorthand for the
presence of these nonrenormalizable quark and gluon interactions
in the effective action. One can therefore compute the couplings
appearing in (\ref{3}) at large scales (say $k=10\ {\rm GeV}$)
in a perturbative QCD calculation of effective vertices like
(\ref{A}).

Due to the presence of quark fluctuations the mass terms $m^2_\Phi$
and $m^2_\chi$ decrease as $k$ is lowered. Perturbation
theory breaks down once
$m^2_\Phi$ or $m^2_\chi$ are of the same size as $k^2$. It
is our central postulate that the mass terms turn negative in the
nonperturbative region of small $k$, inducing nonzero expectation
values\footnote{The $SU(3)_c$ color breaking by nonzero $<\chi>$
occurs only in a gauge-fixed version, whereas an explicitly gauge-invariant
formulation is used below.} for both $<\phi>$ and  $<\chi>$
even for $j=0$.

In an intermediate range of $k$ below $\approx$ 700 MeV the flow of
$m^2_\Phi$ has been discussed previously in a linear quark-meson model
\cite{6}. This is related to the present picture by
neglecting the gluons and $\chi$. In the linear quark-meson
model chiral symmetry breaking was indeed observed due to $m^2_\Phi$
turning negative for small $k$. At least for small enough $k$
the effects of confinement and therefore the gluons should play an
important role, however. They are included here by adding $\chi$ and $A_\mu$.

We suggest that the essential features of confinement are described by
nonzero $<\chi>$ in the range of very small $k$. In the
absence of quark masses we propose
\be\label{x2}
<\chi_{ij,ab}>=\frac{1}{\sqrt{24}}\chi_0\lambda^z_{ji}\lambda^z_{ab}
=\frac{1}{\sqrt6}\chi_0(\delta_{ia}\delta_{jb}-\frac{1}{3}
\delta_{ij}\delta_{ab})\ee
with real $\chi_0>0$. This expectation value is invariant under vectorlike
$SU(3)$ transformations by which identical
left and right flavor rotations and transposed color
rotations are performed with opposite angles. With respect to
the transposed color rotations the quarks behave as antitriplets.
In consequence, under the combined
transformation they  transform as an octet plus a singlet,
with all quantum numbers identical to the baryon octet/singlet. In
particular, the electric charges, as given by the generator $Q=
\frac{1}{2}\lambda_3+\frac{1}{2\sqrt3}\lambda_8$, are integer.
We therefore identify the quark field with the lowest baryon
octet and singlet -- this is quark-baryon duality. Similarly,
the gluons transform as an octet of vector-mesons, again with
the standard charges for the $\rho, K^*$ and $\omega/\varphi$ mesons.
We therefore describe these vector mesons by the gluon field --
this is gluon-meson duality.

In order
to visualize the physical content in this regime, it is convenient
to use  nonlinear coordinates in field space, given by a
hermitean matrix $S$ and unitary matrices $U, v$
\bear\label{4}
\Psi_L&=&U^\dagger N_L v\quad,\quad \Psi_R=N_Rv\quad,\quad
A_\mu=v^TV_\mu v^*-\frac{i}{g}\partial_\mu
v^Tv^*,\nonumber\\
\Phi&=&SU\quad,\quad\chi_{ij,ab}=v^T_{ik}X_{kl,ac}v^*_{lj}U_{cb}
\quad.\quad 
\ear
We also adopt a matrix notation for the color indices
with the quark/baryon nonet represented by a complex $3\times 3$ matrix
$\Psi\equiv\Psi_{ai},\bar\Psi\equiv\bar\Psi_{ia}$.
Besides $v$ all nonlinear
fields are color singlets.
For the present investigation we omit the scalar
excitations except for the Goldstone bosons contained in
$U$. We can therefore replace $X_{kl,ab}$ by the expectation
value (\ref{x2}) and use $<S>=\sigma_0$ such that
\be\label{x3}
\Phi=\sigma_0U\quad,\quad
\chi_{ij,ab}=\frac{1}{\sqrt6}\chi_0\{v_{ai}(v^\dagger U)_{jb}
-\frac{1}{3}U_{ab}\delta_{ij}\}\ee
In terms of the nonlinear field coordinates the Lagrangian
(\ref{2}) reads
\bear\label{5}
{\cal L}&=&Tr\{i\bar N\gamma^\mu\partial_\mu N+i\bar N_L
\gamma^\mu U\partial_\mu U^\dagger N_L+g\bar N\gamma^\mu NV^T_\mu
\}\nonumber\\
&&+\frac{1}{2}Tr\{V^{\mu\nu}V_{\mu\nu}\}
+g^2\chi^2_0\ Tr\ \{V^\mu V_\mu\}-\nu\sigma^3_0\cos\theta\nonumber\\
&&+(
h\sigma_0-\frac{\tilde h}{3\sqrt6}\chi_0)
\ Tr\{\bar N \gamma_5N\}+\frac{\tilde h}{\sqrt6}\ Tr\ \bar N\gamma^5\ Tr\  
N\nonumber\\
&&+(\sigma^2_0+\frac{4}{9}\chi_0^2)\ Tr\ \partial^\mu U^\dagger\partial_\mu  
U+ig\chi^2_0\ Tr\ \{U\partial_\mu U^\dagger V^T_\mu\}\ear
with
$V_{\mu\nu}=\partial_\mu V_\nu-\partial_\nu V_\mu \ -ig[V_\mu,V
_\nu]$.
As it
should be, ${\cal L}$ does not depend on $v$, and $U$ only appears
in derivative terms except for the phase $\theta$ associated
to the $\eta'$-meson. Here we associate $U$ in the standard
way with the pseudoscalar octet of Goldstone bosons
\be\label{7}
U=\exp(-\frac{i}{3}\theta)\exp\left(i\frac{\Pi^z\lambda_z}{f}\right
)\ee
where the decay constant reads
\be\label{8}
f=2(\sigma_0^2+\frac{4}{9}\chi^2_0)^{1/2}\approx 90\ {\rm MeV}.\ee
The nucleon-, the baryon singlet-, and the $\rho$-mass are given by
\be\label{9}
m_N=h\sigma_0-\frac{\tilde h}{3\sqrt6}\chi_0\approx
930\ {\rm MeV}\ ,\ m_1=h\sigma_0+\frac{8}{3}\frac{\tilde h}{\sqrt6}\chi_0\ ,\ 
M_\rho=g\chi_0\approx 770\ {\rm MeV}.\ee
Denoting the relative size of the octet and the singlet
condensates by $x=4\chi^2_0/(9\sigma_0^2)$, one finds for the gauge
coupling
\be\label{10}
g=\frac{4}{3}\left(\frac{1+x}{x}\right)^{1/2}\frac{M_\rho}{f}
\approx 11 \left(\frac{1+x}{x}\right)^{1/2}.\ee
We can compare this with the $(\rho\pi\pi)$-coupling contained
in (\ref{5})
\be\label{11}
-g_{\rho\pi\pi}=g\frac{\chi_0^2}{f^2}=\frac{M^2_\rho}{gf^2}=
\frac{9x}{16(1+x)}g\approx 6\left(\frac{x}{1+x}\right)^{1/2}.\ee
Inferring  $g_{\rho\pi\pi}\approx 6$ from the decay width
$\rho\to2\pi$ (and keeping in mind corrections up to 30 \% from
nonzero quark masses), the agreement between (\ref{10}) and
(\ref{11}) is striking for $x$ not too small. This implies $g\approx12$.
In the gluon language our result corresponds to a finite strong
gauge coupling $\alpha_s\approx12$. One also finds  the
effective coupling of nucleons to the $\rho$-vector mesons
$g_{\rho NN}=g$
.. The
pseudoscalar self-interactions take the standard form
implied by chiral perturbation theory.
The pion nucleon coupling in (\ref{5}) is not
yet in its definite form since the left- and right-handed
nucleon fields can still be redefined by a common (non-integer) power
of $U$. Finally, the $\eta'$-mass
is $M^2_{\eta'}=3\nu\sigma_0/2$. Within a large range of potential parameters 
the source term (\ref{3a}) leads to
realistic masses of the pseudoscalar mesons consistent with
chiral perturbation theory. Electromagnetic and weak interactions
are included by gauging the appropriate flavor symmetries.

We conclude that the simple effective action (\ref{3}) can give
a realistic approximate description of the masses of all low-lying
mesons and baryons and of their interactions. Gluon-meson duality
turns out to be the well-known Higgs phenomenon with colored
scalar fields! The effective action (\ref{3}) is only an approximation,
but its phenomenological success hints to the possibility that
higher-order (``nonrenomalizable'') operators may only play a
subleading role. Higher mass meson and baryon resonances are considered as  
bound states
of the fields used here. At this stage we did not attempt to calculate
the parameters appearing in the potential 
or the Yukawa couplings $h, \tilde h$ from the known short-distance
physics. The
non-perturbative flow equation \cite{5} for the average action seems
to be the appropriate tool for this purpose. Only after a
calculation of $\sigma_0$ and $\chi_0$ in terms of $\Lambda_{\rm QCD}$
we can decide if the proposed field content and the ``truncation''
to a renormalizable effective action are sufficient to solve the
confinement problem quantitatively.\footnote{It is an important
advantage of our setting that the explicit chiral symmetry breaking
by quark mass terms appears only through a linear source term.
The computation of ${\cal L}$ is independent of the quark masses
and the results hold for arbitrary $j$.}

The present framework opens new perspectives for a calculation
of the properties of hadronic matter at high temperature and density.
Especially for high density the issue of baryons vs. quarks plays
a crucial role due to the different Fermi surfaces \cite{7}.
Quark-baryon duality allows for a simple approach to this problem
by using only one field. Recently, a picture of color-flavor-locking
has been proposed \cite{4} for dense baryonic or quark matter.
This ressembles our picture of the QCD-vacuum in many respects.
If true, the question of a phase transition for high density hadronic
matter at zero temperature reduces to the issue of spontaneous
breaking of baryon symmetry: One may expect a second-order phase transition
in the universality class of superfluid He$^4$ if baryon number is
broken in the high density region, and a continuous behavior otherwise.

Quark-baryon duality has profound implications. The nonrelativistic
quark model where baryons are bound states of three quarks is
now supplemented by the view of baryons as ``dressed quarks''. In
a high energy scattering process ``physical quarks'' will come out,
but they will come out with the quantum numbers and masses of
baryons. In this sense, quarks are not confined particles, despite
the fact that color charges remain exactly confined and cannot
appear connected with free particles. This view may have important
consequences for our picture of the parton model, both for structure
functions at small $Q^2$ and for fragmentation. In a first crude
approach to fragmentation one may simply treat the perturbative
quarks as particles with the appropriate masses of the baryons,
associating the quantum numbers in the octet to color and flavor of
the perturbative quarks according to (\ref{4}) with $v=U=1$.
The same holds for the gluons which are now regulated in the infrared
by the vector boson masses. In a second step one has to include
the production of pseudoscalar mesons. Obviously, many new questions
and problems open up with this view. We hope that besides its
conceptual simplicity gluon-meson and quark-hadron duality also
opens new perspectives for quantitative calculations.

\end{document}